\documentstyle[11pt,doublespace,epsf]{article}
\def\picwidth{\epsfxsize=7.0in}        

\def\met{{\not\!\!E_T}}

\begin{document}

\begin{center}
\large{\bf The Charge Asymmetry in $W$-Boson Decays Produced in \\
            $p\overline{p}$ Collisions at \mbox{$\sqrt{s}=1.8$~TeV}}
\end{center}
\vspace{-0.5cm}
\begin{center}
\begin{small}
\hfilneg
\begin{singlespace}
\font\eightit=cmti8
\def\r#1{\ignorespaces $^{#1}$}
\hfilneg
\begin{sloppypar}
\noindent
F.~Abe,\r {13} M.~G.~Albrow,\r 7 D.~Amidei,\r {16} J.~Antos,\r {28}
C.~Anway-Wiese,\r 4 G.~Apollinari,\r {26} H.~Areti,\r 7 M.~Atac,\r 7
P.~Auchincloss,\r {25} F.~Azfar,\r {21}
% ...
\end{sloppypar}

\vskip .025in
\begin{center}
(CDF Collaboration)
\end{center}

\vskip .025in
\begin{center}
\r 1  {\eightit Argonne National Laboratory, Argonne, Illinois 60439} \\
\r 2  {\eightit Istituto Nazionale di Fisica Nucleare, University of Bologna,
I-40126 Bologna, Italy} \\
\r 3  {\eightit Brandeis University, Waltham, Massachusetts 02254} \\
\r 4  {\eightit University of California at Los Angeles, Los
Angeles, California  90024} \\
\r 5  {\eightit University of Chicago, Chicago, Illinois 60637} \\
\r 6  {\eightit Duke University, Durham, North Carolina  27708} \\
\r 7  {\eightit Fermi National Accelerator Laboratory, Batavia, Illinois
60510} \\
\r 8  {\eightit Laboratori Nazionali di Frascati, Istituto Nazionale di Fisica
               Nucleare, I-00044 Frascati, Italy} \\
\r 9  {\eightit Harvard University, Cambridge, Massachusetts 02138} \\
\r {10} {\eightit University of Illinois, Urbana, Illinois 61801} \\
\r {11} {\eightit Institute of Particle Physics, McGill University, Montreal
H3A 2T8, and University of Toronto,\\ Toronto M5S 1A7, Canada} \\
\r {12} {\eightit The Johns Hopkins University, Baltimore, Maryland 21218} \\
\r {13} {\eightit National Laboratory for High Energy Physics (KEK), Tsukuba,
Ibaraki 305, Japan} \\
\r {14} {\eightit Lawrence Berkeley Laboratory, Berkeley, California 94720} \\
\r {15} {\eightit Massachusetts Institute of Technology, Cambridge,
Massachusetts  02139} \\
\r {16} {\eightit University of Michigan, Ann Arbor, Michigan 48109} \\
\r {17} {\eightit Michigan State University, East Lansing, Michigan  48824} \\
\r {18} {\eightit University of New Mexico, Albuquerque, New Mexico 87131} \\
\r {19} {\eightit Osaka City University, Osaka 588, Japan} \\
\r {20} {\eightit Universita di Padova, Instituto Nazionale di Fisica
          Nucleare, Sezione di Padova, I-35131 Padova, Italy} \\
\r {21} {\eightit University of Pennsylvania, Philadelphia,
        Pennsylvania 19104} \\
\r {22} {\eightit University of Pittsburgh, Pittsburgh, Pennsylvania 15260} \\
\r {23} {\eightit Istituto Nazionale di Fisica Nucleare, University and Scuola
               Normale Superiore of Pisa, I-56100 Pisa, Italy} \\
\r {24} {\eightit Purdue University, West Lafayette, Indiana 47907} \\
\r {25} {\eightit University of Rochester, Rochester, New York 14627} \\
\r {26} {\eightit Rockefeller University, New York, New York 10021} \\
\r {27} {\eightit Rutgers University, Piscataway, New Jersey 08854} \\
\r {28} {\eightit Academia Sinica, Taiwan 11529, Republic of China} \\
\r {29} {\eightit Superconducting Super Collider Laboratory, Dallas,
Texas 75237} \\
\r {30} {\eightit Texas A\&M University, College Station, Texas 77843} \\
\r {31} {\eightit University of Tsukuba, Tsukuba, Ibaraki 305, Japan} \\
\r {32} {\eightit Tufts University, Medford, Massachusetts 02155} \\
\r {33} {\eightit University of Wisconsin, Madison, Wisconsin 53706} \\
\r {34} {\eightit Yale University, New Haven, Connecticut 06511} \\
\end{center}
\end{singlespace}
\end{small}
\end{center}
%


\begin{abstract}
The charge asymmetry has been measured using $19,039~W$
decays recorded by the CDF detector during the
1992-93 run of the Tevatron Collider.
The asymmetry is sensitive to the ratio of $d$ and $u$ quark
distributions to $x<0.01$ at $Q^2 \approx M_W^2$,
where nonperturbative effects are minimal.
It is found that of the two current sets
of parton distributions, those of Martin,
Roberts and Stirling (MRS) are favored over the sets most recently
produced by the CTEQ
collaboration.
The $W$ asymmetry data provide a
stronger constraints on $d/u$ ratio than
the recent measurements of $F_2^{\mu n}/F_2^{\mu p}$ which
are limited by uncertainties originating from deutron corrections.
\end{abstract}
\begin{center}
PACS numbers:  13.85.Qk, 13.38.+c, 14.80.Er \\
(Revised version, submitted to PRL.)
\end{center}
\baselineskip=10mm
\vspace{0.3cm}


{}~~~The previous study of the $W$ asymmetry performed using the CDF 1988-89
data~\cite{Old_ASYM}, with less than a quarter of the $\approx 20$~pb$^{-1}$
available for the current analysis,
indicated the potential for hadron collider data to
contribute to our understanding of parton distribution functions (PDFs).
Typically these distributions are extracted from deep inelastic scattering
(DIS) data.  These DIS experiments measure cross sections for electron, muon
or neutrino scattering off nucleon and nuclear targets over some
range of $x$ and $Q^2$.
The PDFs are extracted by fitting these data, within the framework of
perturbative QCD, for the momentum distributions of the proton's constituent
quarks and gluons.  These functions are evolved to high $Q^2$ and
used as input to virtually
every hadronic cross section calculation.  At CDF this fact implies that
uncertainties in the PDFs translate into uncertainties in everything from
a top-quark cross section to a $W$-boson mass measurement; therefore it is
imperative that these distributions are well determined.
 In particular the ratio of $d$ and $u$ quark distributions
is usually extracted from data on the ratio of electron and
muon scattering from neutrons and protons.  Such data suffer from
uncertainties in corrections due to deuteron binding effects~\cite{SHADOW}
and also
from unknown higher twist and nonperturbative effects~\cite{non_pert} at
low values of $Q^2$.  This letter describes
new data which significantly constrain the $u$ and $d$ quark momentum
distributions in the nucleon.

$W^+$ ($W^-$) bosons are produced in $p\overline{p}$ collisions primarily by
the annihilation of $u$ ($d$) quarks from the proton  and $\overline{d}$
($\overline{u}$) quarks from the antiproton.  As the $u$ quark tends to
carry a larger fraction of the proton's momentum than the $d$ quark, the $W^+$
($W^-$) is boosted, on average, in the proton (antiproton) direction.
The charge asymmetry in the production of $W$s, as a function of rapidity
($y_W$), is
therefore related to the difference in the $u$ and $d$ quark distributions
at very
high $Q^2$ ($\approx M^2_W$) and low $x$ ($0.007<x<0.24$) for
$\sqrt{s}=1.8$~TeV and $-1.8<y_W<1.8$.

The $W$ decay involves a neutrino, whose longitudinal momentum is undetermined.
Therefore the quantity measured is the charge asymmetry of the decay leptons,
which has an added contribution due to the $V$-$A$ decay of the $W$.
This portion of the asymmetry has been well measured by muon decay
experiments~\cite{TRIUMF};
thus in comparisons to theory, one can attribute any deviations (between
prediction and measurement) to the PDFs
used in the calculations.  The asymmetry is defined as:
\begin{equation}
A(y_l)=\frac{d\sigma^+/dy_l-d\sigma^-/dy_l}
            {d\sigma^+/dy_l+d\sigma^-/dy_l}
\end{equation}
where $d\sigma^+$ ($d\sigma^-$) is the cross section for
$W^+$~($W^-$) decays to leptons as a function of lepton rapidity ($y_l$),
with positive
rapidity being defined in the proton beam direction. As long as the
acceptance and efficiencies for detecting $l^+$ and $l^-$ are equal, this
ratio of cross sections becomes simply the difference in the number of
$l^+$ and $l^-$ over the sum; all efficiencies and the acceptance as well as
the luminosity cancel.
Further, by CP invariance, the asymmetry at positive
$y_l$ is equal in magnitude and opposite in sign to that at negative
$y_l$, so the two values are combined reducing the
effect of any differences in the efficiencies for $l^+$ and $l^-$.

The CDF detector is described in detail elsewhere~\cite{CDF}.
$W$-boson decays to leptons are identified by the presence of a large amount
of missing transverse energy (${\not\!\!E_T}$)~\cite{met_def} accompanied by
a track in the
central tracking chamber (CTC) which points at either hits in the muon
chambers or a cluster of energy in the electromagnetic (EM) calorimeters.
The CTC is an 84 layer drift chamber which is immersed in a 1.4 T axial
magnetic field.  This magnetic field enables lepton charge determination,
from the curvature of the track, to a high degree of certainty.
Electron candidates are required to fall within the fiducial regions of
either the central, $|y|<1.1$, or the plug, $1.1<|y|<2.4$,
EM calorimeters and to pass identification cuts based on the EM shower's
profile determined with test beam electrons.
Muon candidates are required to have a track in the muon
tracking system, in addition to a minimum ionizing particle
signal in the hadronic and EM calorimeters traversed by the muon track.
The curvature ($C$) of the track is required to be well measured,
$C/\delta C>2$, and  the track must pass within
2~mm of the beam line to reject cosmic rays as well as poorly measured tracks.
Events are required to have a well defined vertex
within 60 cm of the
center of the detector, and ${\not\!\!E_T}>25$ GeV (in the case
of muons after correcting for the muon's momentum).  The transverse energy
($E_T$) of the lepton is required to be greater than 25 GeV.
To reduce the backgrounds due to misidentified dijets, events with a
jet~\cite{jet} whose $E_T$ exceeds 20 GeV are rejected.
The limiting factor in $y$ for this measurement is the rapidity coverage
provided by the CTC.
The data are divided into three
samples: central electrons, plug electrons and central muons.

The triggers for the central electron and muon data sets are checked for
any charge or $E_T$ dependence using data from independent
triggers.  No evidence of such dependencies is found.  The plug electron
triggers, while not having any charge dependence, are not fully efficient at
25~GeV.  Therefore, a correction is determined on a bin by bin basis, using
a Monte Carlo calculation and the measured trigger efficiency, and applied to
the plug electron data.
The correction to $A(y_l)$ is found to be less than 0.005.

Sources of a charge bias in the event selection are investigated by selecting
high $E_T$ electrons or muons, either from a sample of $Z$s or a
sample of $W$s, which satisfy tight kinematic constraints.  No charge
dependent effects are observed.
For example~\cite{My_thesis}, none of the
648 Central-Central $Z$s or 332 Central-Plug $Z$s have same sign leptons,
implying an upper limit on the probability of misidentifing the lepton's
charge of 0.48\% and 0.9\% in the central and plug regions respectively at the
90\% confidence level (C.L.).

The backgrounds to the data (described below) are all typically small.
  In the plug electron sample, misidentified dijet events are the largest
source of background.
This background source is charge symmetric, so it acts to dilute the charge
asymmetry.
The largest background in the central electron sample is due to
$W^{\pm} \rightarrow \tau^{\pm} \nu \rightarrow e^{\pm} \nu\nu\nu$.
For the central muon sample the largest background is  misidentified
$Z\rightarrow \mu^+\mu^-$ where one of the muons is lost out the end of the
CTC.
Misidentified $Z$ decays to electrons are negligible
because the plug and forward calorimeters have a much larger
geometric acceptance than do the muon chambers or the CTC.
The $Z\rightarrow \tau^+ \tau^-$ contamination is
also considered and found to be negligible in all three data sets.
These vector boson related backgrounds are estimated using a Monte Carlo
and detector simulation, and their charge asymmetries are likewise determined.
The cosmic ray contamination of the muon data is negligible.
The $A(y_l)$ values (shown in Fig. 1)
are then corrected on a bin by bin basis for the backgrounds
listed in Table~\ref{bg_percent}, taking into account the shape of
each background's charge asymmetry~\cite{My_thesis}.
The overall systematic uncertainty is very small (as shown in Fig. 2).

\begin{figure}[tbh]
    \begin{center}
    \picwidth
    \mbox{\epsffile[54 283 558 567]{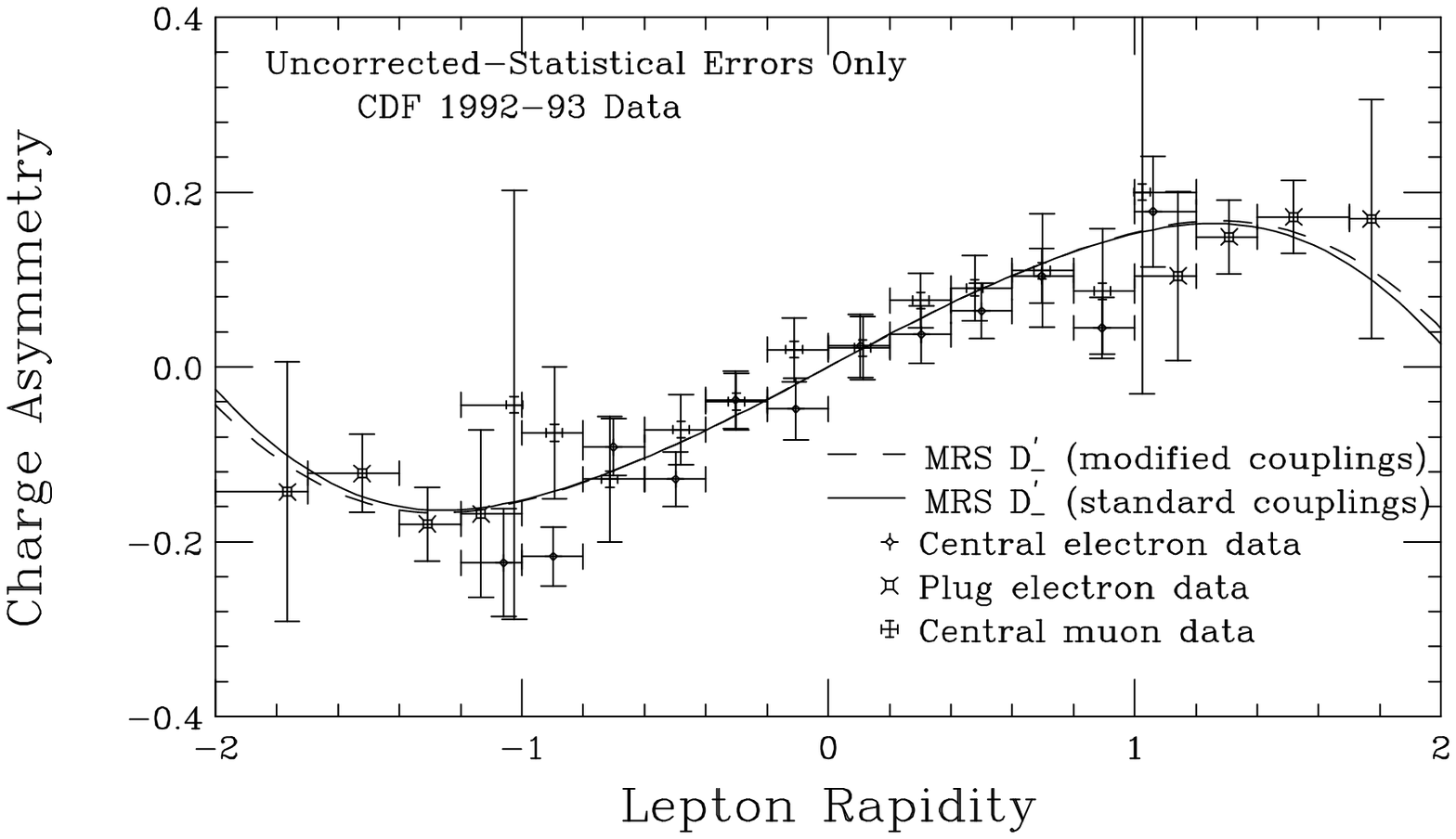}}
    \caption{The charge asymmetry before applying any corrections found
             in each of the detector types (Central EM, Plug EM and Central
	     Muon).  Also shown (dashed line) is the effect of allowing the
	     $W$ couplings to go to their 90\% C.L. limits.}
    \label{asym_unfold}
    \end{center}
\end{figure}

Fig.~\ref{asym_unfold} shows the uncorrected asymmetry before the
values at  positive $y$ are combined with the opposite asymmetry at
negative $y$.
The level of agreement between the various detector types also indicates
that systematic effects are indeed small.
Also shown is the next-to-leading order (NLO) asymmetry
predictions~\cite{Dyrad} made assuming standard $W$ left and right handed
couplings and that which is found when the couplings are allowed to go
to their 90\% C.L. limits~\cite{PRL_V-A}; both calculations use the
MRS~D$_-^\prime$ PDFs as input.
Clearly the uncertainty in the $W$ couplings is much smaller than the
statistical error of the measurement.

\begin{figure}[t]
\begin{center}
\picwidth
\mbox{\epsffile[54 211 567 590]{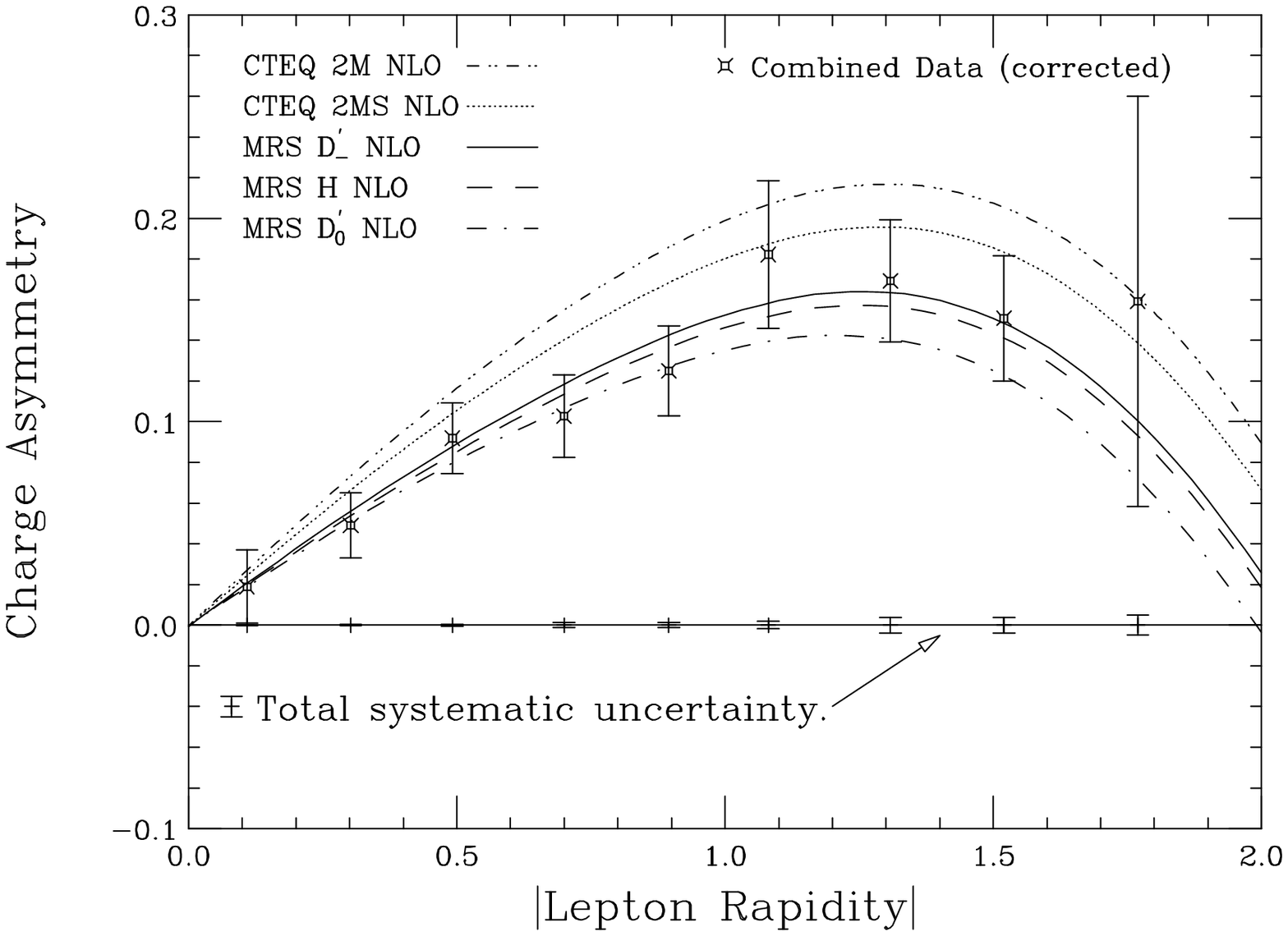}}
\caption{The fully corrected charge asymmetry after the data from the
various detectors
are combined and folded about $y=0$.  The error bars along the
x-axis show the total systematic errors associated with each bin.}
\label{ASYM_PRL}
\end{center}
\end{figure}

Fig.~\ref{ASYM_PRL} shows the fully corrected asymmetry after taking the
weighted mean of the various data sets and the $\pm y$ bins.  The data
are listed in Table~\ref{asym} along with the total uncertainty as well as
the average $y_l$ of the leptons which contribute to each bin.
Also shown are the NLO calculations~\cite{Dyrad} made using several sets
of parton distributions~\cite{PDF} as input.
{\it The $A(y_l)$ measurement was not included in any of these
PDF determinations,
therefore it provides an independent test of the PDFs.}
To quantify the degree to which the various PDFs
reproduce the data, Table~\ref{chi} lists the results of $\chi^2$ tests of the
goodness of fit.  There is no differentiating power in
the first and last $y$ bins. In particular, the last bin
is statistically limited because the $W$ production cross-section
is small at large $y$.
Therefore the $\chi^2$ is calculated for the
seven bins spanning $0.2<|y|<1.7$ and for the weighted mean of the bins (the
theoretically calculated
asymmetries were weighted in the identical manner).  The motivation for the
last test is that all the modern PDFs predict asymmetries with
essentially the same shape and only differ in overall magnitude.

As can be seen in Table~\ref{chi}, our data exclude the
older MRS E$^\prime$, MRS B$^\prime$
and MT B1 distributions, which were extracted before the recent precision,
high statistics DIS data were available.
What is more significant is the extent to which the asymmetry data
favor the recent MRS distributions (MRS~D$_0^\prime$, MRS~D$_-^\prime$
and MRS~H) over the most recent CTEQ2 distributions, as both groups had
access to the same recent DIS data.

  The $W$ charge asymmetry is particularly sensitive to the slope of the $d/u$
ratio versus $x$~\cite{slope1,slope2}, whereas the $F_2^{\mu n}/F_2^{\mu p}$
measurements are sensitive to the magnitude of this ratio as well as
to the quantity $\overline{u}~-~\overline{d}$.
Recently NMC has measured $F_2^{\mu n}/F_2^{\mu p}$~\cite{NMC} over an $x$
range comparable to that accessible at CDF (though at much lower $Q^2$).
The NMC data~\cite{F2NP,SHADOW}
 were used to constrain $d/u$ in the most recent parton
distribution fits.
For easier comparison of the $d/u$ slopes,
Figure 3b shows the $d/u$ ratios after being shifted by a constant
so they agree with MRS~D$_0^\prime$ at $x=0.2$.
The distributions which predict the largest average slope of the $d/u$
ratio over the $x$ range $0.007-0.24$, also predict the
largest charge asymmetry.
One sees that even though the MRS and CTEQ PDFs
were both determined by fitting to the $F_2^{\mu n}/F_2^{\mu p}$ data,
they
have very different
$d/u$ distributions and thus very different charge asymmetry predictions.
This is because $F_2^{\mu n}/F_2^{\mu p}$ is also sensitive to the
differences in the $\overline{u}$ and $\overline{d}$ distributions,
whereas the $A(y_l)$ asymmetry is not as sensitive.
CTEQ's parameterization of the $\overline{u}$ and $\overline{d}$ sea
distributions compensates for a steep $d/u$ ratio~\cite{MRS_CTEQ} and
leads to a prediction
for $F_2^{\mu n}/F_2^{\mu p}$ which is consistent with the NMC data
but is much less consistent with the $A(y_l)$ measurement presented in this
paper.

In summary, the $W$ charge asymmetry measurement from CDF is showing
sensitivity to the slope of the $d/u$ quark distribution at
a level of precision which is already better than deep inelastic scattering
experiments, which have additional uncertainties originating from
unknown higher twist and nonperturbative
effects~\cite{non_pert} at low values of $Q^2$ at small $x$, and also
uncertainties in the extraction of neutron cross sections from deuterium
data~\cite{SHADOW}.
    The uncertainty in the slope of the $d/u$ quark distribution is
    the dominant contribution to the systematic error from PDFs
    in the extraction of the $W$ mass from collider data.
    These new asymmetry measurements already can be used to substantially
    reduce the errors on the $W$ mass~\cite{MRS_W}.
The upcoming run, with its four fold increase in integrated luminosity,
promises to cut the uncertainties in half, as the $A(y_l)$ systematic
errors are small.

     We thank the Fermilab staff and the technical staffs of the
participating institutions for their vital contributions.  This work was
supported by the U.S. Department of Energy and National Science Foundation;
the Italian Istituto Nazionale di Fisica Nucleare; the Ministry of Education,
Science and Culture of Japan; the Natural Sciences and Engineering Research
Council of Canada; the National Science Council of the Republic of China;
the A. P. Sloan Foundation; and the Alexander von Humboldt-Stiftung.

\newpage
%
\begin{table}
\begin{center}
\caption{Backgrounds (\%) in the $W\rightarrow e\nu$ and
$W\rightarrow \mu \nu$ charge asymmetry event samples.
The values in boldface were
used to correct the measurement in conjunction with the background's charge
asymmetry.}
\vspace{0.25in}
\begin{tabular}{cccc}  \hline \hline
Source & Central $e$ & Plug $e$ & Central $\mu$ \\
\hline
$W\rightarrow \tau \nu$ &$\bf{2.0\pm0.2}$&$\bf{2.0\pm0.2}$ &$\bf{2.0\pm0.2}$ \\
  QCD                   &$\bf{0.4\pm0.1}$&$\bf{4.1\pm0.9}$ &$\bf{0.3\pm0.1}$ \\
 $Z\rightarrow ee~or~\mu\mu$ & $< 0.2$  & $< 0.2$         &$\bf{4.7\pm0.7}$ \\
$Z\rightarrow \tau\tau$ & $< 0.1$  & $< 0.1$  & $< 0.1$    \\ \hline \hline
\end{tabular}
\label{bg_percent}
\end{center}
\end{table}
\begin{table}
\begin{center}
\caption{The charge asymmetries (after all corrections) and total uncertainties
           in the combined $e$ and $\mu$ channels.}
\vspace{0.25in}
\begin{tabular}{cccccc}  \hline \hline
$|y_l|$ bin & $\langle|y_l|\rangle$ & $A(y_l)$ & $\sigma$  \\ \hline
0.0-0.2 & 0.11 & 0.019 & $\pm0.018$ \\
0.2-0.4 & 0.30 & 0.049 & $\pm0.016$ \\
0.4-0.6 & 0.49 & 0.092 & $\pm0.017$ \\
0.6-0.8 & 0.70 & 0.103 & $\pm0.020$ \\
0.8-1.0 & 0.90 & 0.125 & $\pm0.022$ \\
1.0-1.2 & 1.08 & 0.182 & $\pm0.036$ \\
1.2-1.4 & 1.31 & 0.169 & $\pm0.030$ \\
1.4-1.7 & 1.52 & 0.151 & $\pm0.031$ \\
1.7-2.0 & 1.77 & 0.16 & $\pm0.10$ \\
\hline \hline
\end{tabular}
\label{asym}
\end{center}
\end{table}
\begin{table}
\begin{center}
\caption{The $\chi^2$ comparisons between
the predicted asymmetries (calculated at NLO) for several NLO PDFs including
the most recent MRS and CTEQ distributions.  The comparison of the weighted
means, $\overline{A}(y_l)$ indicates
the MRS H distributions fit the asymmetry data best.
The very recent PDFs ( CTEQ3 and MRS A) are not included in the comparison,
since the CDF asymmetry data was included in these fits.}
\vspace{0.25in}
\begin{tabular}{ccccc}  \hline \hline
     & \multicolumn{2}{c}{$0.2<|y_l|<1.7$}
     & \multicolumn{2}{c}{$\overline{A}(y_l)$}  \\
PDF Set & $\chi^2$ (7 $dof$)& ${\cal P}(\chi^2)$& $\Delta\sigma$
     & ${\cal P}(\sigma^2)$ \\ \hline
   CTEQ 2M        &24.&$<0.01$& 4.6& $<0.01$ \\
   CTEQ 2MS       &11.& 0.15 & 2.9& $<0.01$ \\
   CTEQ 2MF       &17.& 0.02 & 3.8& $<0.01$ \\
   CTEQ 2ML       &15.& 0.04 & 3.5& $<0.01$ \\ \hline
   CTEQ 1M        &6.1& 0.52 & 2.1& 0.04 \\
   CTEQ 1MS       &3.9& 0.79 & 1.5& 0.13 \\
   MT B1          &17.& 0.02 &-3.2& $<0.01$ \\ \hline \hline
   MRS H          &1.8& 0.97 &-0.1& 0.96 \\
MRS D$^{\prime}_-$&1.9& 0.97 & 0.5& 0.61 \\
MRS D$^{\prime}_0$&3.6& 0.83 &-0.9& 0.35 \\ \hline
   HMRS B         &4.2& 0.75 &-1.2& 0.23 \\
    KMRS B$_0$    &19.& 0.01 &-3.6& $<0.01$ \\
MRS E$^{\prime}$  &30.& $<0.01$ &-4.9& $<0.01$ \\
MRS B$^{\prime}$  &24.& $<0.01$ &-4.1& $<0.01$ \\ \hline \hline
    GRV NLO       &12.& 0.12 & 3.0& $<0.01$ \\ \hline \hline
\end{tabular}
\label{chi}
\end{center}
\end{table}
\newpage
%
\begin{figure}
\begin{center}
\picwidth
\mbox{\epsffile[90 153 540 630]{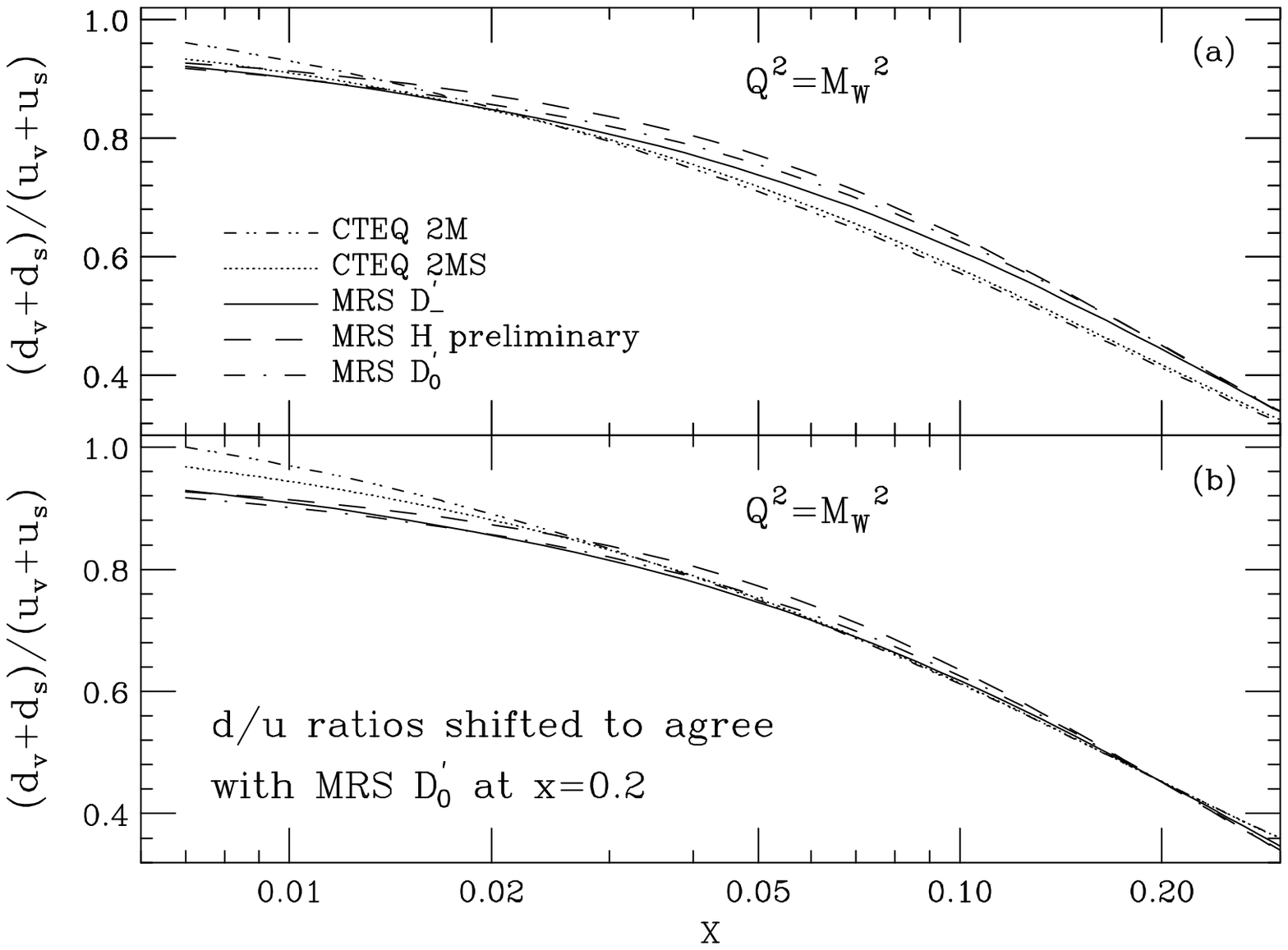}}
\caption{
(a) The $d/u$ ratios for various parton distributions.
(b) The $d/u$ ratios of various PDFs after they have been shifted to agree
with MRS D$_0^{\prime}$ at x=0.2; those which have the largest average slope
predict the largest asymmetry.}
\label{MRS_CTEQ_F2}
\end{center}
\end{figure}


\begin{thebibliography}{999}

\bibitem{Old_ASYM}
F. Abe {\em et al}., Phys. Rev. Lett. {\bf 68}, 1458 (1992).
%
\bibitem{SHADOW}
B. Badelek, J. Kwiecinski, Nucl. Phys. B {\bf 370}, 278 (1992);
 Note that these deuteron shadowing corrections are model dependent and it
 is not clear whether they should be applied to the data.
%
\bibitem{non_pert}
L. Whitlow {\em et al.}, Phys. Lett. B {\bf 282}, 475 (1992);
M. Arneodo {\em et al.}, Phys. Lett. B {\bf 309}, 222 (1993);
M. Virchaux and M. Milsztajn, Phys. Lett. B {\bf 274}, 221 (1992).
%
\bibitem{TRIUMF}
B. Blake {\em et al.,} (TRIUMF Coll.), Phys. Rev. D {\bf 37}, 587 (1988).
%
\bibitem{CDF}
F. Abe {\em et al}., (CDF Coll.), Nucl. Instrum. Methods
Phys. Res., Sect. A {\bf 271}, 387 (1988).
%
\bibitem{met_def}
Transverse energy, $E_T$, is defined as $E\times sin~\theta$, where $\theta$
is the
polar angle of the energy cluster with respect to the proton beam direction.
Missing $E_T$, \mbox{$\met$}, is defined as the negative magnitude of the
vector sum  of $E_T$ in all calorimeter cells with $|\eta|<3.6$.
%
\bibitem{jet}
The jet energy is clustered in a cone of radius 0.7 in $\eta$-$\phi$ space.
%
%
\bibitem{My_thesis}
M. Dickson, University of Rochester thesis (May 1994), UR-1349 (unpublished).
%
\bibitem{Dyrad}
W. Giele, E. Glover and D.A. Kosower,
Nucl. Phys. {\bf B403}, 633 (1993).
%
\bibitem{PRL_V-A}
Particle Data Group, Phys. Rev. D {\bf 45}, VI.16 (1992).  The left and right
vector couplings are set to $|g^V_{RL}|=0.11$ and $|g^V_{LL}|=0.96$.
%
%
\bibitem{PDF}
%
J. Botts {\em et al.} (CTEQ Coll.),
Phys. Lett. B {\bf 304}, 159 (1993).\\
%
%
%
%
A.D. Martin, R.G. Roberts and W.J. Stirling, (MRS Coll.),
RAL-93-077 (1993).
%
%
%
%
\bibitem{slope1}
E. L. Berger {\em et al.}, Phys. Rev.
D {\bf 40}, 83 (1989).
%
\bibitem{slope2}
A.D. Martin, R.G. Roberts and W.J. Stirling,
Mod. Phys. Lett. A {\bf 4}, 1135 (1989).
%
\bibitem{NMC}
P. Amaudruz {\em et al}., (NMC Coll.),
Phys. Lett. B {\bf 295}, 159 (1992);
Nucl. Phys. B {\bf 371}, 3 (1992);
M. Arneodo {\em et al.}, CERN-PPE-93-117(1993).
%
\bibitem{F2NP}
A.D. Martin, R.G. Roberts and W.J. Stirling,
Phys. Lett. B {\bf 306}, 146 (1993).
%
\bibitem{MRS_CTEQ}
A.D. Martin, R.G. Roberts and W.J. Stirling,
RAL-94-055, DTP/94/34 (June 1994).
%
\bibitem{MRS_W}
A.D. Martin and W.J. Stirling,
Phys. Lett. B {\bf 237}, 551 (1990).
%
\end{thebibliography}
\end{document}